\documentclass{article}
\usepackage{spconf,amsmath,graphicx}
\usepackage{hyperref}

\title{Improving Prosody for Cross-Speaker Style Transfer by Semi-Supervised Style Extractor and Hierarchical Modeling in Speech Synthesis}

\name{Chunyu Qiang$^*$, Peng Yang$^*$, Hao Che, Ying Zhang, Xiaorui Wang, Zhongyuan Wang}

\address{ Kwai, Beijing, P.R. China}

\begin{document}
%
\maketitle
\begin{abstract}

  Cross-speaker style transfer in speech synthesis aims at transferring a style from source speaker to synthesized speech of a target speaker's timbre. In most previous methods, the synthesized fine-grained prosody features often represent the source speaker's average style, similar to the one-to-many problem(i.e., multiple prosody variations correspond to the same text).  In response to this problem, a strength-controlled semi-supervised style extractor is proposed to disentangle the style from content and timbre, improving the representation and interpretability of the global style embedding, which can alleviate the one-to-many mapping and data imbalance problems in prosody prediction. A hierarchical prosody predictor is proposed to improve prosody modeling. We find that better style transfer can be achieved by using the source speaker’s prosody features that are easily predicted. Additionally, a speaker-transfer-wise cycle consistency loss is proposed to assist the model in learning unseen style-timbre combinations during the training phase.  Experimental results show that the method outperforms the baseline. We provide a website with audio samples \href{https://qiangchunyu.github.io/style-transfer/STW.html}{$^1$}.

\end{abstract}

\renewcommand{\thefootnote}{\fnsymbol{footnote}} 
\footnotetext[1]{Equal Contribution.} 
\footnotetext[2]{Audio samples: https://qiangchunyu.github.io/style-transfer/STW.html}

\section{Introduction}
\label{sec:intro}

With the development of deep learning, speech synthesis technology has rapidly advanced\cite{wang2017tacotron,li2019neural,elias2021parallel}. Improving the expressiveness and controllability of TTS systems for a better listening experience has attracted more attention and research. So far, cross-speaker style transfer TTS is divided into two categories: global style transfer \cite{wu2021cross, skerry2018towards, li2022cross,wang2018style,zhang2019learning, habib2019semi} and fine-grained prosody transfer \cite{lee2021styler, pan2021cross, yi2022prosodyspeech,chen2022fine}.

Many global style transfer methods using style-id as a global style variable have been proposed\cite{wu2021cross, skerry2018towards, li2022cross}. There are correlations between style-ids such as happy and surprised, and the distribution of emotions in complicated datasets is complex and varied. There is a one-to-many problem with using style-id to describe emotions since it is impossible to guarantee that data with the same style ID consistency in the intensity of emotion, such as generally sad, very sad, and extremely sad. Reference encoder methods based on global style tokens(GST)\cite{wang2018style} or variational autoencoders (VAEs)\cite{zhang2019learning, habib2019semi, kenter2019chive, qiang2022style} are widely used to learn the latent representation of style state in a continuous space. VAE is used to model the variance information in the latent space with Gaussian prior as a regularization. The so-called "speaker leakage problem" arises when synthetic speech appears to have been uttered by the source speaker rather than the target speaker due to the fact that the style being transferred came from speech uttered by the source speaker. Many methods use intercross training, gradient reversal, domain adversarial training or add multiple loss functions\cite{lee2021styler,zhang2021denoispeech, zhou2021seen, xue2021cycle, an2021improving, joo2020effective} to reduce the source speaker leakage. The speaking styles are characterized by localized prosody variations, many fine-grained prosody transfer methods using both global style variable and local prosody variable have been proposed\cite{lee2021styler, pan2021cross, yi2022prosodyspeech,chen2022fine}.  
Most of the previous methods use the source style-id, text and target speaker-id to predict style prosody features. The synthesized fine-grained prosody features often represent the average style of source speaker. In practice, the multi-style data of the source speaker is sparse, and the data of target speaker only contains single-style data without labels. This makes it difficult to predict the target speaker's prosody features (style of the source speaker). Meanwhile, the phone-level prosody features are distorted, making predictions inaccurate. The contributions of this paper include: 

\begin{figure*}[t]
 \centering
 \includegraphics[width=\linewidth]{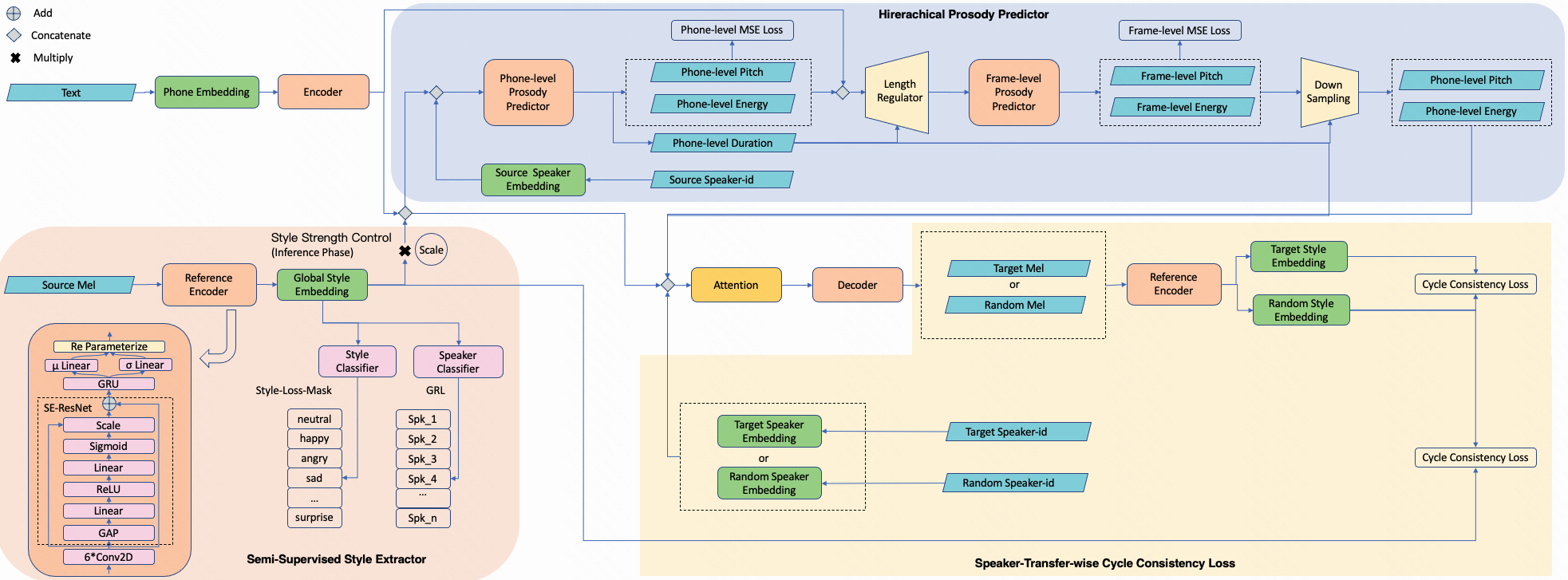}
 \caption{The architecture of proposed model.}
 \label{fig:proposed_model}
\end{figure*}

\begin{itemize}
\item A strength-controlled semi-supervised style extractor is proposed to disentangle the style from content and timbre, improving the representation and interpretability of the global style embedding, which can alleviate the one-to-many mapping and data imbalance problems in prosody prediction.

\item A hierarchical prosody features predictor is proposed to improve prosody modeling. The phone level prosody features are distorted (lack of information relative to the frame level features) leading to prediction difficulties. However, we expect that local style variables only contribute only to the information at the phone-level while more personalities within the phone are learned through target speaker-id and global style embedding. We find that better style transfer can be achieved by using the source speaker’s prosody features that are easily predicted.

\item A speaker-transfer-wise cycle consistency loss is proposed to assist the model in learning unseen style-timbre combinations during the training phase in order to address the instability and speaker leakage problem produced by the source speech and predicted source prosody features. 

\end{itemize}

\section{Method}
The proposed framework is illustrated in Fig.1. As shown, the proposed model is an attention-based seq2seq framework, hierarchical prosody features predictor take a text sequence, a source speaker-id and a global style embedding as input to predict source speaker's phone-level prosody features. Tacotron-like systems take a text sequence, a target speaker-id, a global style embedding and predicted source prosody features as input, and use autogressive decoder to predict a sequence of acoustic features frame by frame. 

\begin{figure*}[t]
 \centering
 \includegraphics[width=\textwidth]{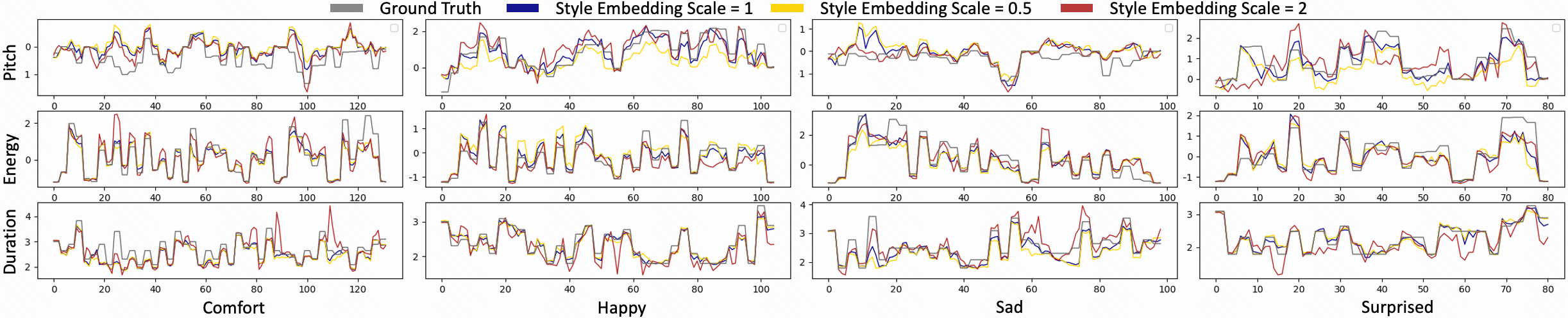}
 \caption{Prosody features predicted by scaling global style embedding(The abscissa represents the phoneme length).}
 \label{fig:speech_production}
\end{figure*}

\subsection{Semi-Supervised Style Extractor}

\subsubsection{Reference encoder} \label{VAE}

As illustrated in Fig.1, in order to alleviate the highly entangled problem in cross-speaker style transfer and improve the style extraction ability of the model, a style bottleneck sub-network\cite{pan2021cross} is introduced to the reference encoder. The style bottleneck network consists of 6 layers 2D convolutional networks and a (Squeeze-and-Excitation based ResNet architecture) SE-ResNet block \cite{hu2018squeeze}. The SE-ResNet block can adaptively recalibrate channel-wise feature responses by explicitly modelling interdependencies among channels, and produce significant performance improvements.  The model obtains a continuous and complete latent space distribution of styles through the VAE \cite{kingma2013auto} structure to improve the style control ability. A 64-dimensional vector is sampled from Gaussian distribution as global style embedding. Random operations in the network cannot be processed by backpropagation, "reparameterization trick" is introduced to VAE: $\boldsymbol{z} = {\hat{\mu}} + {\hat{\sigma}} \odot \phi ; \phi \sim \mathcal{N}(0, I)$. Three tricks are used to solve KL collapse problem: 1) The KL annealing is introduced. 2) A staged optimization method is adopted to optimize the reconstruction loss first and then the KL loss. 3) A margin $\Delta$ is introduced to limit the minimum value of the kl loss as shown:${L}_{kl} = max(0, D_{KL}[\mathcal{N}({\hat{\mu}},{\hat{\sigma}}^2)||\mathcal{N}(0, I)]-\Delta)$. Furthermore, the style strength of synthesized speech can be effectively controlled by scaling global style embedding.

\subsubsection{Style Loss Mask} \label{SST}
The speaker timbre and style in speech signals are highly entangled, and reducing the source speaker leakage plays an important role in the task of cross-speaker style transfer. The model uses a gradient reversal layer(GRL) for adversarial speaker training. The extracted global style embedding is fed into the speaker classifier, which consists of a fully connected layer, a softmax layer and a GRL. To improve the representation and interpretability of the global style embedding, we add a style classifier that is consistent with the speaker classifier structure. Since the majority of the target speaker data lacks style labels and labeled multi-style data is sparse, categorizing unlabeled data as neutral will affect the style classifier's accuracy and decrease the global style embedding's capacity for representation. For semi-supervised training, we mask the style classification loss of such target speakers in each batch to zero, and the reference encoder will subtly identify the styles in each speech contains. 

\subsection{Hierarchical Prosody Predictor}
The phone level prosody features are distorted (lack of information relative to the frame level features) leading to prediction difficulties. However, we expect that local style variables only contribute only to the information at the phone-level while more personalities within the phone are learned through target speaker-id and global style embedding. As shown in the Fig.1, a hierarchical prosody features predictor is proposed to improve the accuracy of phone-level features prediction. In order to reduce the speaker information of the prosody features and improve the stability, the extracted prosody features are standardized by the mean variance at the speaker level. We find that better style transfer can be achieved by using the source speaker’s prosody features that are easily predicted. The phone embedding, source speaker embedding and global style embedding are fed into the phone-level prosody predictor to obtain the pitch, energy and duration of phone-level. Phone-level pitch and energy features are concatenated with phone embedding, expanded using a length regulator, and used as input to frame-level prosody predictor. The predicted frame-level pitch and energy features are downsampled by calculating the mean of each phoneme to obtain the final phone-level prosody feature. Both the phone-level and frame-level prosody feature predictors consist of 2 layers 1D convolutional networks and one layer fully connected network. To ensure that the length of the frame-level prosody features is consistent with the ground truth to calculate the frame-level mean square error (MSE) loss, the duration of ground truth is used for expansion in the training phase. 

\subsection{Speaker-Transfer-wise Cycle Consistency Loss}
In the training phase, the combination of target speaker embedding and source style embedding as input is an out-of-set problem, since there is no ground truth to calculate the reconstruction loss. Most of the existing methods use the ground truth acoustic features and the synthesized acoustic features constitute paired two-tuples to compute cycle consistency loss. Due to teacher-forcing, these two features are almost the same, making this method less effective. As shown in the Fig.1, a speaker-transfer-wise cycle consistency loss is proposed. The randomly chosen speaker-id and the target speaker-id are used as input to calculate the forward twice in each training step, and the rest of the input information is entirely consistent. We expect that the target mel-spectrogram and random mel-spectrogram will have different timbres and the same style. Two cycle consistency losses are constructed: (random style embedding \& target style embedding), (random style embedding \& global style embedding). The method assist the model in learning unseen style-timbre combinations during the training phase in order to address the instability and speaker leakage problem produced by the source speech and predicted source prosody features.

\section{Experiments}

\subsection{Experimental Step}
A dataset is used with 20 native Mandarin speakers (10 males and 10 females), two of which contained multi-styles (comfort, happy, sad, surprised, natural), while the others contained only single style (similar to natural) without style labels. Each labeled multi-style speaker has 300 sentences per style. Each unlabeled single-style speaker has 10,000 sentences. The dataset has an average per-speaker duration of 2.9 seconds, and all speech waveforms sampled at 24kHz are converted to mel-spectrogram with a frame size of 960 and hop size of 240.  In the inference phase, the centroid of the global style embeddings extracted from all 
sentences for each style is used. The front-end model structure is consistent with \cite{qiang2022back}. The vocoder used in this experiment is LPCNet \cite{valin2019lpcnet}.

\subsection{Compared Models}

To our best knowledge, {\bf Disentangling}\cite{li2022cross}  and {\bf  Bottleneck}\cite{pan2021cross} are two state-of-the-art strategies that are used in the speech style transfer task. Here, to show the superiority of our proposed method, these two strategies are also adopted to compare with our method. To be fair, we changed all models to make use of the same attention-based seq2seq framework. The phone-level prosody predictor structure of proposed model and  {\bf Bottleneck} is the same. An ablation study is performed by comparing the proposed method with several variants achieved by removing style loss mask ({\bf SLM}) method (described in Sec 2.1.2.) or speaker-transfer-wise ({\bf STW}) cycle consistency loss(described in Sec 2.3.). 

\subsection{Test Metrics}
All the subjective tests are conducted by 11 native judgers, and each metrics consisted of 20 sentences per style. The test metrics used in the evaluation are listed below:

\begin{itemize}
\item {\bf Prosody Measurement}: Phone-level prosody correlation to source style recording, include pitch(F0), Duration and Energy.

\item {\bf Strength Perception}: A subjective strength perception test. The judger is asked to sort them according to the style strength (weak, medium, and strong).

\item {\bf Style and Speaker Similarity MOS}: To verify similarity in expected speaking style and timbre between source speech and synthesized speech. 

\item {\bf Style Perception}: A subjective style perception test. The judger is asked to select one from 5 options (comfort, happy, sad, surprised, neutral), according to his/her perception on the test case.
\end{itemize}

\begin{table}[]
 \caption{Prosody Measurement}
 \label{tab:prosody}
 \centering
\begin{tabular}{llll}
\hline
Model              & F0            & Energy        & Duration      \\ \hline
Bottleneck\cite{pan2021cross} & 0.59          & 0.88          & \textbf{0.87} \\ \hline
Proposed           & \textbf{0.70} & \textbf{0.92} & 0.86          \\ \hline
\end{tabular}
\end{table}

\begin{table}[]
 \caption{Strength Perception Accuracy}
 \label{tab:strength}
 \centering
\begin{tabular}{lllll}
\hline
Model                 & Comfort        & Happy          & Sad            & Surprised      \\ \hline
Disentangling\cite{li2022cross}  & 39.1          & 53.64          & 50.00          & 56.36          \\ \hline
Proposed              & \textbf{66.36} & \textbf{70.00} & \textbf{72.73} & \textbf{73.63} \\ \hline
\end{tabular}
\end{table}

\subsection{Results}
The prosody measurements in Table 1 ({\bf Disentangling} does not support fine-grained prosody prediction) show that the proposed hierarchical prosody predictor is significantly better than single-level model {\bf Bottleneck}. The frame-level loss provides more detailed undistorted supervision, and the prediction results in the pitch and energy are closer to the ground truth. As shown in Table 2 ({\bf Bottleneck} does not support strength control), the proposed method achieves better style strength control due to fine-grained prosody features. Unlike \cite{li2022cross}, which does not care about the ordering direction, only samples arranged in a weak(scale=0.5)-medium(scale=1)-strong(scale=2) order are treated as correct. As shown in Fig.2, the synthesized phone-level prosody features of each synthesis are plotted based on different style embedding scales. As can be seen, in each subfigure, the features trajectories of different strengths present a similar trend but with different values. For instance, the pitch reduces and the duration increases as the scale increases from 0.5 to 2 for comfort and sad. As for happy and surprised, the pitch increases and the duration reduces as the scale increases, and the result was as expectd. The significant effect of our proposed method on adjusting the style strength is demonstrated. 
\begin{table}[]
 \caption{MOS}
 \label{tab:mos}
 \centering
\begin{tabular}{lll}
\hline
Model                 & Style Sim & Speaker Sim                                     \\ \hline
Disentangling & 3.57 ± 0.091         & 3.89  ± 0.082 \\ \hline
Bottleneck     & 3.81 ± 0.080              & 4.01 ± 0.072                                                     \\ \hline
Proposed(w/o SLM)     & 3.23 ± 0.044           & 3.94   ±  0.016                                                  \\ \hline
Proposed(w/o STW)     & 3.92 ± 0.042               & 3.98   ±  0.033                                                  \\ \hline
Proposed              &  \textbf{3.99  ±  0.082}                &        \textbf{4.02   ±   0.077}                                          \\ \hline
\end{tabular}
\end{table}

\begin{table}[]
 \caption{Style Perception Accuracy}
 \label{tab:style}
 \centering
\begin{tabular}{lllll}
\hline
Model                 & Comfort     & Happy       & Sad         & Surprised   \\ \hline
Disentangling & 47.27          & 45.45          & 54.55          & 45.45          \\ \hline
Bottleneck    & 65.45          & \textbf{57.27} & \textbf{78.18} & 54.55          \\ \hline
Proposed(w/o SLM)     & 43.64          & 34.54          & 43.64          & 32.73          \\ \hline
Proposed(w/o STW)     & 70.91        & 56.36 & 76.36      &    \textbf{56.36}         \\ \hline
Proposed              & \textbf{72.73} & 54.55          & \textbf{78.18} & \textbf{56.36} \\ \hline
\end{tabular}
\end{table}

As shown in Table 3, in terms of speaker similarity MOS, both methods have achieved acceptable results. The proposed method achieves similar scores to {\bf Bottleneck}. In terms of style similarity MOS, compare with {\bf Proposed(w/o SLM)}, {\bf SLM} method gives the model more explicit style information, the proposed method achieves a best style similarity. The subjective test for style evaluation are shown in Table 4, the proposed method achieves the best performance, where the style loss mask method is effective for the style representation ability. Compared with {\bf Proposed(w/o STW)}, the improvement of {\bf STW} method is weak. During the experiment, we find that the initial value of the cycle consistency loss is very low, indicating that the global style embedding already has a good style representation ability, and does not contain the source speaker's timbre information. The {\bf STW} method may be more efficient when using the traditional reference encoder structure. We will verify this hypothesis in future.

\section{Conclusions}

In this paper, we focus on the prosody prediction in the cross-speaker style transfer task. A strength-controlled semi-supervised style extractor, a hierarchical prosody features predictor, and a speaker-transfer-wise cycle consistency loss are proposed. We achieve good style transfer by using the source speaker’s prosody features. Experiments show that the effectiveness of our proposed methods. 
\vfill\pagebreak

\bibliographystyle{IEEEbib}

\small
\bibliography{strings,refs}

\end{document}